\documentclass{article}

\usepackage{arxiv}

\usepackage[utf8]{inputenc} 
\usepackage[T1]{fontenc}    
\usepackage[hidelinks]{hyperref}      
\usepackage{url}            
\usepackage{booktabs}       
\usepackage{amsfonts}       
\usepackage{amsmath}
\usepackage{nicefrac}       
\usepackage{microtype}      
\usepackage{lipsum}		    
\usepackage{graphicx}
\usepackage{natbib}
\usepackage{doi}
\usepackage{rotating}
\usepackage{makecell}
\usepackage{xcolor}
\usepackage[normalem]{ulem}
\usepackage{soul}
\usepackage{longtable}
\usepackage{array} 
\usepackage{booktabs}
\usepackage{enumitem} 
\usepackage{comment}
\usepackage{caption}
\usepackage{adjustbox}
\usepackage{floatrow}

\graphicspath{ {./Figures/} }

\newcommand{\pkg}[1]{{\normalfont\fontseries{b}\selectfont #1}}
\let\proglang=\textsf
\let\code=\texttt

\usepackage{floatrow}
\usepackage{caption}
\usepackage{subcaption}
\usepackage[rightcaption]{sidecap}
\usepackage{longtable,booktabs,array}

\title{Exploring the Shape of Economics: \\ A Multilayer Network Analysis of Social Communities and Intellectual Similarity Among Journals Before and After the 2008 Financial Crisis}

\date{} 					

\author{\href{https://orcid.org/0000-0003-0293-482X}{\includegraphics[scale=0.06]{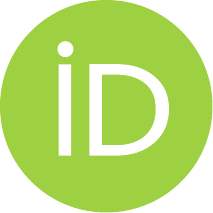}\hspace{1mm}Alberto Baccini}
  \\
	Dipartimento di Economia Politica e Statistica\\
	Università degli Studi di Siena\\
	Siena, Italy \\
	\texttt{alberto.baccini@unisi.it} \\
	\And \href{https://orcid.org/0000-0002-7210-4541}{\includegraphics[scale=0.06]{orcid.pdf}\hspace{1mm}}{Lucio Barabesi}\\
	Dipartimento di Economia Politica e Statistica\\
	Università degli Studi di Siena\\
	Siena, Italy \\
	\texttt{lucio.barabesi@unisi.it} \\
	\And \href{https://orcid.org/0000-0001-5231-4763}{\includegraphics[scale=0.06]{orcid.pdf}\hspace{1mm}}{Carlo Debernardi}\\
	Dipartimento di Economia Politica e Statistica\\
	Università degli Studi di Siena\\
	Siena, Italy \\
	\texttt{carlo.debernardi@unisi.it} \\
	}

\renewcommand{\headeright}{}
\renewcommand{\undertitle}{}
\renewcommand{\shorttitle}{Exploring the Shape of Economics} 

\hypersetup{
pdftitle={},
}

\makeatletter
\input{size12.clo}
\makeatother

\begin{document}

\maketitle

\begin{abstract}

This paper develops a multilayer network approach for exploring the evolution of scientific disciplines, using the case of economics before and after the 2008 global financial crisis as a large-scale empirical testing ground. The units of analysis are journals, linked by social and intellectual relationships. The analysis covers all journals indexed in EconLit across three years (2006, 2012 and 2019). In the most recent year (2019), the dataset includes 909 journals, over 30,000 editorial board members, more than 260,000 authors, 134,000 articles, and nearly 2 million cited references. For each period, we model journals as connected in a four-layer multiplex network: the social relationships are based on shared editors (interlocking editorship) and shared authors (interlocking authorship), while the intellectual ones are based on shared references (bibliographic coupling) and textual similarity between articles.  These four layers are integrated using Similarity Network Fusion to produce unified similarity networks from which journal communities are identified. Comparing the field across the three periods reveals a high degree of structural continuity. Although research topics changed after the crisis, the fundamental social and intellectual relationships among journals remained remarkably stable. A major result of the analysis is that editorial networks play the dominant role in shaping hierarchies and legitimize knowledge production within the discipline. Whether this finding holds in other scientific disciplines remains an open question for future research.

\end{abstract}

\keywords{Economics journals, bibliographic coupling, text-embedding, gatekeepers of science, interlocking editorship. \textbf{ECONLIT}: A1: General Economics, B2: 	History of Economic Thought since 1925}

\section{Introduction}

Scientific journals are the primary outlets for contemporary research across nearly all disciplines \citep{Lariviere}. A general consensus has emerged about their traditional function: they certify, assign priority, disseminate, and archive scientific knowledge \citep{rowland, Priem}. Yet, despite their centrality, we know surprisingly little about how journals function as social and intellectual systems, how they are connected through shared editors, authors, citations, and textual content, and how these connections evolve in response to major external shocks. This paper develops a multilayer network approach to investigate these questions. We adopt a dual perspective, examining both social relationships and intellectual proximity. Economics provides an ideal case study: following the 2008 global financial crisis, the discipline experienced public scrutiny over its assumptions, models, and ability to foresee systemic risks. We leverage this crisis as a large-scale empirical testing ground to investigate whether and how scientific fields change over time.

The crisis has indeed become a central reference point for discussions of the state of economics as an academic discipline. The famous question posed by Queen Elizabeth II at the London School of Economics in spring 2009 — ``Why did no one see it coming?'' — is the starting point for a discussion of the realism of existing macroeconomic models and the collective ability of the profession to foresee systemic risks. The British Academy responded to Queen's question by attributing the crisis to ``a failure of the collective imagination of many bright people … to understand the risks to the system as a whole''. However, this view was sharply contested by a group of British economists who argued that the failure was not merely one of imagination but also the result of economics becoming detached from real-world complexities and sustained by unrealistic assumptions and uncritical views of markets \citep{dow2009}. \citet{Colander} even argued that the financial crisis itself was rooted in ``the systematic failure of the economics profession'', pointing to deep structural flaws in the discipline. \citet{mirowski} reached a similar conclusion, emphasizing how the dominant intellectual framework had rendered economists blind to systemic risks. Together, these critiques frame the crisis not only as a policy or market failure but as evidence of a profound epistemic breakdown within economics itself. Beyond the specific case of economics, however, this crisis provides a case study to investigate a broader question of interest to the science of science: how do scientific disciplines respond to major external shocks?

More precisely, in the aftermath of the crisis, three main interpretations emerged of its effects on the discipline of economics. One acknowledges significant gaps in macroeconomic theory revealed by the crisis, but argues that these can be addressed through targeted refinements of existing models rather than a wholesale paradigm shift \citep[see, among the others,][]{caballero, blanchard}. The economics profession was ``slow to see'' but ``fast to act'', with a rapid collective shift toward studying its causes and consequences \citep{LEVY2022}. A second interpretation views the crisis as a moment of reckoning, exposing the limitations of dominant models and prompting calls for methodological pluralism, greater historical awareness and stronger engagement with neighboring social sciences \citep{Colander, Fourcade}. A third view contends that, despite the criticism, the core structures of academic economics remained largely unchanged, with only minor adjustments in research agendas and teaching practices \citep{mirowski}. 

These debates resonate with a long-standing theme in the history of economic thought: the idea that major external shocks have often reshaped the discipline. The Great Depression is related to the rise of the Keynesian revolution in macroeconomics \citep[for all][p. 9]{marchionatti2}, while the stagflation of the 1970s is associated with the crisis of Keynesianism and the emergence of new classical and supply-side approaches \citep[again, for all,][p. 415]{marchionatti3}. In this light, the 2008 crisis is often interpreted as another critical juncture in the evolving relationship between economic theory, empirical evidence, and policy practice \citep{duarte, mirowski}. 

The contrasting interpretations of the effects of the financial crisis on the discipline of economics intersect with a broader body of literature that examines its internal organization, its boundaries, and its evolving specialties. Scholars have explored how economics interacts with other disciplines and how its internal differentiation unfolds over time. Umbrella terms such as ``economics imperialism'' \citep{Lazear}, ``reverse imperialism'' \citep{reverse, ambrosino_2023, heisse}, ``specialization'' \citep{Cedrini}, ``empirical turn'', ``fragmentation'', ``mainstream pluralism'' and heterodoxy \cite[for a review, see][]{dippoliti}, have been used to capture different aspects of this evolution and to situate debates about the discipline’s response to the crisis within longer-term trajectories.

These phenomena are most often explored through qualitative, close-reading approaches that trace the evolution of economic ideas and the development of economics' intellectual traditions over time. Such perspectives highlight how conceptual frameworks, methodological debates, and shifting boundaries between economics and adjacent sciences have shaped the evolution of the discipline. Alongside this tradition, a smaller body of empirical research — reviewed in the second section — has employed bibliometric techniques and distant-reading methods. Taken together, the bulk of this literature, both qualitative and bibliometric, has primarily concentrated on intellectual and thematic developments, while the social structures underpinning the discipline have been considered only marginally.

This paper develops a multilayer network approach to address this gap. It focuses on economics journals as the basic unit of analysis. Journals represent institutionalized spaces where both social networks and cognitive content are made visible and traceable over time. 

From a social point of view, two key communities are observable around each journal: editorial board members and authors who have published in it. These two communities have different characteristics. Editorial boards are structured differently across journals, but in almost all cases, they follow a hierarchical organization. Typically, one or more scholars serve as editor-in-chief, supported by associate editors and regular board members \citep{baccini_re}. The board members act as gatekeepers of scientific communication by managing peer review and making publication decisions \citep{crane1967}. These decisions also determine the inclusion or exclusion of scholars from the ``social circle'' of the authors of a journal \citep[p. 13]{Craneinvis}. 
Unlike the editorial board, the social circle of authors of a journal is not hierarchically ordered ``there is no formal leadership [...] although there are usually central figures'' \citep[p.14]{Craneinvis}. 

From an intellectual or cognitive standpoint, a journal can be conceptualized as a curated collection of scientific articles. Each article contributes with its contents to the evolving knowledge base of the field. The reference lists within these articles provide insight into the epistemic foundations and intellectual lines that define the thematic orientation of the journal. Textual analysis and citation analysis serve thus as a proxy for the cognitive structure of the discipline, capturing how ideas circulate, cluster, and transform over time.

We explore the four layers of information represented by editors, authors, texts, and references to get a representation of the social and intellectual relational structures that connect journals. We model journals as connected in a four-layer multiplex network. Editors and authors are used for building two interlocking social structures in which journals are linked if they share at least one editor or author. Two types of intellectual relationships are also examined. First, article abstracts are used to measure the similarity of content between pairs of journals via text-embedding techniques. Second, references are used to construct bibliometric coupling relationships, which can be interpreted as a representation of the knowledge base shared by journals. These four layers are then integrated using Similarity Network Fusion to produce unified similarity networks from which journal communities are identified.

These structural relationships are analyzed for all journals indexed in \textit{Econlit} considering data for a period before the financial crisis (2006) and two periods after (2012 and 2019). This approach allows us to assess the extent to which the crisis reshaped the field, either by reinforcing or disrupting existing disciplinary configurations. 

The paper is organized as follows. The second section presents a review of the relevant literature.The third presents data, methods, and the workflow of the exploratory analysis. The fourth section presents the main results of the quantitative analysis, and the fifth discusses the clustering of journals. The two final sections are devoted to discussion and conclusion, respectively.

\section{Related works}

As anticipated, this section reviews the small body of empirical research that has employed bibliometric techniques and distant reading methods for analyzing intellectual and thematic developments of economics. We do not consider the huge amount of literature aimed at ranking scholars, journals, or institutions active in economics based on bibliometrics, especially citation, data. This stream of literature builds on evaluative bibliometrics, while our contribution adopts a \textit{positive} approach, aimed at describing and explaining phenomena in economics as a discipline.

The most comprehensive effort is that of \citet{Claveau}, who have used bibliographic coupling between articles to map the increasing specialization of economics over recent decades. Their work provides systematic evidence of how the discipline has evolved from a relatively unified field into a mosaic of specialized domains. \citet{Truc_2021} examine developments in economic methodology through co‑citation analysis of articles published in the two main journals in the field over the past three decades. They construct separate co‑citation networks for each decade and compare their structures to evaluate the historical trajectories proposed in existing interpretations. 

Citation data are used as the main evidence in the debate on the intellectual insularity of economics. \citet{angrist} claimed they have documented a ``clear rise in the extramural influence of economic research'' and at the same time an increase in references to other social sciences in the economics literature. These results are definitely challenged by \citet{Truc_2023} who studied since the 1950s in thousands of journals. They show that economics was historically among the most insular disciplines but has grown more open since 1990, particularly to management, environmental sciences, and other social sciences and humanities (SSH). Despite this shift, economics remains one of the least interdisciplinary fields in the SSH domain, alongside management. Moreover, unlike other major social sciences, its leading journals have played little role in fostering interdisciplinarity.

Some exercises using distant reading techniques were also developed, usually based on very limited corpora of economics articles \citep[for example ][]{kosnik, bowles, galiani, bowles2025}. A notable exception is \cite{Ambrosino} who consider the full text of 250 thousands journal articles, They apply LDA topic modeling to trace the evolution of economics over time and to identify structural changes within the discipline. The combined use of citation and textual evidence is proposed by \citet{garcia}, who analyze the subfield of consumption modeling over forty years. Their study integrates co‑citation networks with semantic information derived from frequent word sequences in the abstracts of co‑citing articles, allowing them to observe the joint evolution of citation clusters and thematic content.

Although much of this literature focuses on intellectual and thematic developments, a smaller number of studies have turned their attention to the social structure underlying the discipline. \citet{Claveau} briefly touched on authorship patterns and their implications for the organization of economic research communities. The authorship in a selected set of journals was first explored by \citet{hodgson1999} -- replicated by \citet{Aistleitner} -- who found evidence of an ``institutional oligopoly''. Some papers analyzed the role of research evaluation and especially the use of quantitative indicators for the evolution of contemporary economics \citep{lee, dippoliti, dippoliti_cit, corsi}.

The role of editor of journals was first explored by \citet{hodgson1999} and, in a gender perspective, by \citet{addis2003}. \citet{Colussi} investigated the role of social connections between authors and editors in the publication process of several prominent economics journals. \citet{onder} documented the relation between the diffusion of topics and the appointment of economic editors in the case of \textit{American Economic Review}. \citet{heckman2020} analyzed the role of the top five economic journals in shaping the recruitment and career system in the field. More directly, \cite{baccini2009, baccini2010, baccini_re} examined the network of editorial relationships between economics journals, shedding light on how interlocking editorships and shared social affiliations contribute to the institutional reproduction and coherence of the field. 

With a few exceptions, previous works separately treat the intellectual and social dimensions. Instead, social studies of science, from their inception, have tried to trace the evolution of a discipline as involving not only shifting ideas and topics, but also changes in the social and organizational infrastructure that supports knowledge production. For example, more than 50 years ago, \citet{mullins1972development} suggested that the study of the development of specialties within a discipline should ``consider both intellectual and social variables''. Mullins examined the emergence of molecular biology from traditional biology, arguing that it resulted from a combination of ``normal intellectual and social activities'' of scholars. He defined intellectual activities as paradigm development, problem success and problem solving; while social activities includes communication, co-authorship, collegueship and apprenticeship.

More recently, the so-called ``science of science'' programmatically conceptualized together the intellectual and social dimensions of science. \citet{fortunato} suggested to represent science ``as a complex, self-organizing, and evolving network of scholars, projects, papers, and ideas''. This idea originates from previous scientometric studies mainly devoted to the delineation of scientific fields. At least since \citet{Borgman}, they are are defined by considering three ``theoretical variables'': producers of communication (authors), artifacts of communication (articles), and concepts. Although qualitative studies may handle together the intellectual and social dimensions, quantitative studies usually treat them separately due to the difficulties of developing methodologies for integrating information of deeply different nature. \citet{Zitt} proposed the notion of ``hybridization'' or of ``multi-network approaches'' to label contributions that try to combine different layers of information for scientific fields delineation and classification purposes. A recent review \citep{baccini_sim_net_fus}, collected only a few papers trying to handle together intellectual and social dimensions with quantitative methods.

This paper builds on the hybridization approach. 
In a first study, \citet{baccini_intellectual} explored to what extent intellectual proximity among scholarly journals is also proximity in terms of social communities gathered around the journals. They considered three research fields, among them economics. Intellectual similarity was proxied by the journal co-citation network, while social proximity was computed by considering two social networks: the first generated by interlocking authorship, i.e.\ by authors writing in more than one journal, the second by interlocking editorship, i.e.\ by scholars sitting on the editorial board of more than one journal. The structure of the three networks in terms of dissimilarity have a high correlation, by indicating ``that intellectual proximity is also proximity between authors and between editors of the journals''. Thus, the three maps of editorial power, intellectual proximity, and author communities tell similar stories.

In a subsequent paper, \citet{baccini_sim_net_fus} adopted, for the first time in a scientometric environment, a technique called ``Similarity Network Fusion'' (SNF) \citep{wang2014a}. Similarity relations among journals are represented as a three-layer multiplex of co-citations, common authors and common editors. The information contained in the three layers is combined by constructing a fused similarity network. For economics, the structure of the fused network depends mainly on editors, suggesting that their role as gatekeepers of journals is the most relevant in defining the boundaries of scholarly communities.

In a third paper, \citet{baccini_cje} developed a hybridized measure of similarity at the article level, again using the SNF technique. The relationships between articles published in the \textit{Cambridge Journal of Economics} are organized in a two-layer multiplex with similarities based on the full-text of articles and  bibliographic coupling. SNF permitted to combine the information into a new network. A clustering algorithm is applied to the fused network, and a fine grained classification of articles is obtained. The authors compared also SNF technique with other available techniques of hybridization, claiming that it has ``the main advantage of relieving researchers of the responsibility of choosing weights that can determine the final structure of the integrated network. SNF is based on local properties of the input networks, and the contribution of each starting network to the fusion can be rigorously measured ex-post'' \citep[p. 24]{baccini_cje}. 

In this paper, the SNF is used to explore the change in intellectual and social relationships between economic journals over a period of about 15 years.

\section{Data and methods}

As anticipated, economics journals are the basic unit of our analysis. We consider data on economics journals in three different periods: one (2006) before and two (2012 and 2019) after the 2008 financial crises. For each year and journal, four types of information are collected: 
\begin{enumerate}[noitemsep, nosep]
    \item the list of members of its editorial board;
    \item the list of authors who have published at least one paper in a three year period starting from the reference year;
    \item the abstracts of the articles published in journals in a three year period starting from the reference year;
    \item the cited references of the articles published in journals in a three year period starting from the reference year.
\end{enumerate}
For each journal, editorial board members and authors represent the social communities gathered around it; abstracts can be used to delineate its intellectual content, while cited references represent its knowledge base.

The use of three-year time windows for the bibliographic metadata enables us to gather more information regarding each individual journal, and is arguably justified by the relatively long publication times in Economics: i.e.\ the editorial board in charge in the reference year has reasonably been responsible for most of the publications of the following years, despite potential unobserved changes in its composition.

\subsection{Data collection}

The journals considered for the analysis are those covered by the ``Gatekeepers of Economics Longitudinal Database'' (GOELD). It contains the editorial boards of economics journals, as described in detail in \citet{baccini_re}. Data are collected considering one year each decade from 1866 to 2006 and the two years 2012 and 2019. GOELD will be released when the research project - of which this article is a part - is completed. In this paper, editor names are standardized within each year, and for such a reason it is not possible to precisely calculate a measure of their turn-over.

GOELD includes 1,724 journals indexed in \textit{EconLit}. The list of journals was compiled from the website of the American Economic Association (AEA) in April 2019\footnote{\url{https://web.archive.org/web/20190716024210/https://www.aeaweb.org/econlit/journal_list.php}}. According to \citet{Gusenbauer}, \textit{EconLit} provides the best bibliographic coverage of the economics literature and a limited coverage of the business literature.

To complement the data on the composition of the editorial boards, we also collected the metadata of the journal publications. Our choice as a data source is OpenAlex \cite{openalex}: an open catalog and citational database. From a substantive point of view, this enables us to cover a larger share of non-english and/or nationally focused outlets -- traditionally underrepresented in databases like Scopus or Web of Science.

As previously mentioned, we considered the reference years 2006, 2012 and 2019 (in line with the GOELD database) and collected the metadata for research works published in each reference year, as well as those published in the two subsequent years.

Most journals have been identified through their International Standard Serial Number (ISSN), while 59 have been manually matched since they were present in the OpenAlex database, but their entry was missing an ISSN. The dataset obtained in this way was composed of 552,861 works. However, it was still including non-research related published materials like front and back covers, editorial statements, and reviewers acknowledgements.

\begin{table}[h]
    \centering
    \begin{tabular}{ccccccc}
        \toprule
        Year & Active journals & Covered journals & Editors & Authors & Abstracts & References \\ 
        \midrule
        2006   & 1,269   & 702   & 17,178   & 122,309   & 74,710  & 826,449 \\ 
        2012   & 1,510   & 941   & 25,612   & 178,361   & 105,618  & 1,267,189 \\ 
        2019   & 1,516   & 909   & 30,675   & 260,315   & 134,279  & 1,950,863 \\
        \bottomrule
    \end{tabular}
    \caption{Number of journals, editors, authors, abstracts and references.}
    \label{data}
\end{table}

We thus further reduced the initial sample by keeping only entries categorized as research articles according to both the OpenAlex and Crossref classification and including at least one reference. This resulted in a final sample of 379,956 (\textasciitilde69\%) articles, from which we obtained the authorship and cited references.
Only 314,607 (\textasciitilde83\%) of the works in the final sample had sufficient textual content in the abstract. Of those, 14,203 (\textasciitilde4\%) featured a non-English abstract. Since the adopted text-embedding models were trained only on English sources, we opted to automatically translate these abstracts to English via the DeepL API.

Table \ref{data} reports the number of journals, editors, authors, abstracts, and references used for the exploratory analysis.

\subsection{Similarity matrices}

The building blocks of the analysis consist of suitable representations of the social and intellectual relations between journals. Social relations between pairs of journals are measured (i) as similarities in the composition of their editorial boards and (ii) as similarities in the sets of their authors. Intellectual relations between pairs of journals are measured (i) as similarities between abstracts and (ii) as similarities in their knowledge base by using a bibliographic coupling approach. These four blocks of information are then used to build, for each period, a four-layer multiplex network. Each layer of a multiplex represents one type of information, where journals are nodes and edges are similarities between journals. The four layers give rise to four similarity matrices, which were constructed as follows.

The journal similarity matrix based on editors is computed as the pairwise cosine similarity between rows of the unweighted bipartite matrix, with journals in the rows and editors in the columns. Thus, edges link journals that share editorial board members. The journal similarity matrix based on editors is computed with an analogous procedure, with the only difference that the authors were weighted according to the fractional count of their publications in each journal. The journal similarity matrix based on cited references is computed similarly but weighted by the number of times each reference appears in each journal. Given the unstructured nature of the data, the abstract-based similarity matrix is less straightforward to implement. To obtain such a matrix, we first computed the embedding vector of each abstract using SentenceBERT \citep{reimers2019}. This text embedding approach associates a high-dimensional numerical vector to each text, in such a way that proximity in this abstract space correlates with semantic similarity. We then computed the medoid of each journal, that is the article with the lowest total cosine distance from all the others, and finally defined the similarity matrix as the cosine similarity among the medoids.

The information contained in the four-layer multiplex is then combined together with a technique called Similarity Network Fusion (SNF) \citep{wang2014a}. The result of this procedure is a fifth similarity matrix for each year, which pools information from each underlying dimension. It should be noticed that the first three matrices contain only null or positive entries, while a such guarantee does not hold for the one based on abstract. However, in order to apply the SNF, negative entries in the starting matrices are not allowed. Thus, a suitable transformation is applied, mapping cosine similarities to proper similarities, which are functions of Euclidean distances (see Appendix, Section A.1).

\subsection{Workflow of the exploratory analysis}

As anticipated, for each observed year economics journals and their similarity relationships are represented as a four-layer multiplex network. The first step of exploratory analysis consists of evaluating the degree of association between the information collected in the four layers. To this end, two different analyses are considered: 
(i) the Generalized Distance Correlation (GDC) \citep{szekely2007} is used to evaluate the association between the structure of the matrices. It is defined in the interval $[0,1]$; large distance correlation values between two matrices reflect a strong overlap in the information they encode, while the small values correspond to different informational content;
(ii) the Louvain community detection algorithm is applied to each of the four layers to determine whether journals can be grouped into distinct clusters \citep{Newman2006}. Partitions are evaluated using modularity, a method that quantifies the strength of the community structure within a network. 

As anticipated, for each period, the second step of the exploratory analysis consists of the integration of the information encoded in the four layers using SNF \citep{wang2014a}. Once the fused network is built, the conditional and unconditional dependence of the fused network on the individual layer of information. The unconditional dependence is measured by calculating the GDC between the fused network and each of the layers. Conditional dependence indicates the contribution of each layer to the fused network and is estimated as the Partial Distance Correlation (PDC) \citep{szekely2014} between the fused network and each layer, conditioned on the structure of the other layers (see Appendix, Section A.2). The community structure of the fused network is then explored using the Louvain algorithm. 

The comparison between the three observed periods is carried out by calculating the GDC between the three fused networks. However, this analysis is not technically straightforward: the GDC metric requires networks to share identical node sets, whereas the dynamic nature of journal populations, due to foundation, closure, and merger, results in networks with non-overlapping node sets across the years. To reconcile these differences, we apply a couple of techniques that align the node sets. The details of these methods are reported in the next section. In principle, this comparison may lead to two different scenarios: one in which the structural relationships among journals change over the years, and another in which these relationships remain stable. The occurrence of one scenario or the other is decisive for the subsequent step, which involves defining journal communities on the three fused matrices. In the first scenario, a differentiated classification for each period will be required. However, in the second scenario, a single classification of journals can be adopted over the three considered periods.

As we shall see, the second scenario prevails, with overall stability in the network structures throughout the observed period. The detection of communities of journals for reaching a unique coherent classification over the three periods was a challenge for two main reasons. First, not all journals are observed across the three periods; despite this, we had strong evidence of a stable global structure. Second, the fused matrix is dense, which means that the stochastic part of community detection algorithms might easily determine a large number of nodes to swing across different communities depending on random variations rather than stable structural relationships. To solve these issues, we adopted a relatively simple and intuitive approach called Cluster-based Similarity Partitioning \citep{strehl2002}. We iteratively applied the Louvain community detection algorithm 1,000 times to each of the three fused matrices and then pooled the results. In this way, we obtained a weighted network connecting each of the possible pairs of journals with a strength proportional to the number of times that pair appeared in the same cluster. We then only kept the relations among journals that co-appeared in the same cluster at least 80\% of the times. This produced a highly clustered network, with around 20\% of the journals becoming isolates, that is they do not clearly belong to any community. Finally, we applied the Louvain algorithm one last time in the network obtained in this way. Figure \ref{fig:ensemble_full} reports a graphical representation of the result.

The last step of the exploratory analysis consists in studying the contribution of the four layers of information to the internal structure of each group of journals and to the relationships between groups of journals. In this case also, the techniques exploited are GDC and PDC. GDC and PDC were computed in the \proglang{R}-computing environment \citep{RN13} using the \code{dcor} and \code{pdcor} functions in the package \pkg{energy} \citep{RN29}. Communities and modularities were computed in \proglang{R} by using the function \code{cluster\_louvain} in the package \pkg{igraph} \citep{Csardi}. The similarity network fusion was obtained in \proglang{R} using the function \code{SNF} in the package \pkg{SNFtool} \citep{snf}. Finally, visualization is developed in \proglang{Gephi}.

\section{Data analysis}

As anticipated, the GDC, as reported in Table \ref{GDC_4_layers}, allows us to evaluate the degree of overlap across the four layers of information separately for each observed period. For all three periods, the social layers are highly correlated, indicating that editors and authors generate very similar relations across journals. References are also correlated with the two social layers. The layer of the abstracts shows for the three periods a small value of GDC with the other three layers, indicating that the information gathered by abstracts has a lower overlap with the others. 

\begin{table}
\small
\caption{Generalized distance correlation between similarity matrices 2006, 2012, 2019.}
    \centering
    \begin{tabular*}{\textwidth}{@{\extracolsep\fill}ccccc}
    \toprule %
        \textbf{2006} & Authors & Editors & Abstracts & References \\
        \midrule
        Authors & 1 & 0.984 & 0.473 &  0.857 \\
        Editors & ~ & 1 & 0.471 & 0.853 \\ 
        Abstracts & ~ & ~ & 1 & 0.406 \\ 
        References & ~ & ~ & & 1 \\ 
    \midrule
        \textbf{2012} & &  &  &  \\ 
        \midrule
        Authors & 1 & 0.984 & 0.445 &  0.772 \\
        Editors & ~ & 1 & 0.447 & 0.776 \\ 
        Abstracts & ~ & ~ & 1 & 0.344 \\ 
        References & ~ & ~ & & 1 \\ 
    \midrule%
        \textbf{2019} &  & & & \\ 
        \midrule
        Authors & 1 & 0.948 & 0.429 &  0.700 \\
        Editors & ~ & 1 & 0.449 & 0.713 \\ 
        Abstracts & ~ & ~ & 1 & 0.318 \\ 
        References & ~ & ~ & & 1 \\ 
    \bottomrule%
            \end{tabular*}
    \label{GDC_4_layers}
\end{table}

The results of community detection in the four layers through the Louvain algorithm are reported in Table \ref{Communities_in_layers}. This table reveals persistently low modularity values across all network layers for the three periods considered, signalling weak community structure in each of the layers. Authors and editors show fluctuating community counts, and their modularity remains minimal $<0.009$, reflecting highly interconnected, non-distinct clusters. Abstracts exhibit marginally higher modularity $[0.026, 0.028]$ but still small absolute values, suggesting limited thematic compartmentalization. References show a declining number of communities from $15$ to $10$ with low stable modularity $[0.01, 0.012]$, with a faint indication of an increasingly homogeneous use of references. Collectively, these patterns question the robustness of community detection when the information layers are considered separately. 

\begin{table}[ht]
\caption{Communities and modularities in the four layers and in the fused network for the years 2006, 2012 and 2019.}
\centering
\resizebox{\textwidth}{!}{
\begin{tabular}{lcccccc}
    \toprule
  \ & \multicolumn{2}{c}{\textbf{2006}}  & \multicolumn{2}{c}{\textbf{2012}} & \multicolumn{2}{c}{\textbf{2019}} \\
  & Communities & Modularity & Communities & Modularity & Communities & Modularity \\ 
  \midrule
  Authors & 138 & 0.008  & 162 & 0.006 & 136 & 0.006\\ 
  Editors & 139 & 0.009  & 138 & 0.007 & 139 & 0.007\\ 
  Abstracts & 4 & 0.028  & 5 & 0.026 & 4 & 0.026\\ 
  References & 15 & 0.012 & 10 & 0.01 & 10 & 0.011\\ 
  Fused & 81 & 0.685 & 89 & 0.686 & 89 & 0.686\\ 
\bottomrule
\end{tabular}
}
\label{Communities_in_layers}
\end{table}

The second step of the analysis consists of the pooling of information across layers using SNF, separately for each of the three periods. The three fused matrices are computed by setting the number of neighbors $k = 20$, the hyperparameter $\alpha = 0.5$, and the number of iterations $T = 20$. The application of the Louvain algorithm to detect communities in fused networks results in a rather stable number of communities ($[81, 89]$) with stable high modularity ($[0.685, 0.686]$), indicating that groups of journals emerge when information on social and intellectual similarities is integrated into a single measure. 

Table \ref{gdc_fused_layers} reports the GDC between the fused network and each individual layer. GDC allows evaluating the total dependence of the fused network between the fused network and each individual layer of information, without controlling for other layers. Hence, the GDC reflects the overall structural alignment between the fused network and each individual layer. Authors and editors exhibit near-perfect GDC values across all three periods, indicating that their configurations are almost fully embedded within the fused network. References display a steady decline in association, signaling a gradual decoupling from the fused structure, while Abstracts maintain stable but limited alignment. 

\begin{table}[ht]
\caption{Generalized distance correlation between the fused matrix and each of the layers. Years 2006, 2009, 2019.}
\small
\centering
\begin{tabular}{lccc}
  \toprule
             \textbf{GDC}       & \textbf{2006} &\textbf{2012} &\textbf{2019} \\ 
             \midrule
  Fused-Abstracts   &0.491 & 0.457 & 0.461 \\ 
  Fused-Authors      & 0.993 & 0.989& 0.953\\
  Fused-Editors    & 0.99 & 0.994 &  0.995\\ 
  Fused-References  & 0.854 &  0.768 &  0.712\\ 
   \bottomrule
\end{tabular}
\label{gdc_fused_layers}
\end{table}

The partial distance correlations reported in Table \ref{pdc_fused_layers} verify the independent contribution of each layer to the fused network structure, controlling for the other three layers. In other words, PDC isolates unique contributions of each layer by conditioning on the remaining layers. Editors display the strongest independent influence: PDC is in the interval $[0.336, 0.453]$, with stable values between years. Authors and Abstracts show declining contributions, indicating their diminishing role in shaping the fused structure when conditioned on other layers. References exhibit fluctuations, partially recovering in 2019 but remaining below initial levels. These results underscore Editors as the most consistent driver of the fused network architecture, while other layers display weaker or decreasing conditional contributions over time.

While GDC shows that Authors and Editors dominate the total shared variance with the fused network, PDC highlights Editors as the primary drivers of the structure of the fused networks, independent of other layers, across the years. 

\begin{table}[ht]
\caption{Partial distance correlation between the fused matrix and each of the layers, conditioned to the other three layers. Years 2006, 2009, 2019. }
\small
\centering
\begin{tabular}{lccc}
  \toprule
             \textbf{PDC}       & \textbf{2006} &\textbf{2012} &\textbf{2019} \\ 
             \midrule
  Fused-Abstracts $\mid$ Authors, Editors, References   &0.263 & 0.254 & 0.243 \\ 
  Fused-Authors $\mid$ Abstracts, Editors, References      & 0.297 & 0.21& 0.161\\
  Fused-Editors $\mid$ Abstracts, Authors, References    & 0.335 & 0.455 &  0.419\\ 
  Fused-References $\mid$ Abstracts, Authors, Editors  & 0.336 &  0.262 &  0.307\\ 
   \bottomrule
\end{tabular}
\label{pdc_fused_layers}
\end{table}


The intertemporal comparison of the fused matrices requires that they contain the same nodes. Two different strategies were developed for this purpose. The first consists of considering the GDC among annual matrices containing only journals active in the three years considered (column ``intersection'' of Table \ref{inter-temporal}). The second consists of adding to the annual matrices non-active journals (column ``imputed'' of Table \ref{inter-temporal}). The similarity values of these journals with respect to all other journals are imputed to zero if a journal was not yet active or ceased its activity in the year under consideration. In the few cases where a journal was active in 2006 and 2019, but data for 2012 are missing, similarity values are calculated as the average of the two known values.

\begin{table}[ht]
\caption{Inter-temporal generalized distance correlation between fused matrices. Intersection indicates matrices of the journals present in all the three years considered. Imputed indicates matrices containing all the journals present in at least one year out of the three considered.}
\centering
\small
\begin{tabular}{lcc}
  \toprule
  & Intersection & Imputed \\ 
  \midrule
  2006-2012 & 0.999 & 0.999 \\ 
  2012-2019 & 1 & 0.999 \\ 
  2006-2019 & 0.999 & 0.999 \\ 
   \bottomrule
\end{tabular}
\label{inter-temporal}
\end{table}

The intertemporal GDC between the fused matrices shows a nearly perfect correlation, indicating that the relationship structures between the economics journals do not change between before and after the financial crisis.

\section{Ensemble clustering of economics journals}

\begin{figure}
\centering
     \includegraphics [scale=0.23]{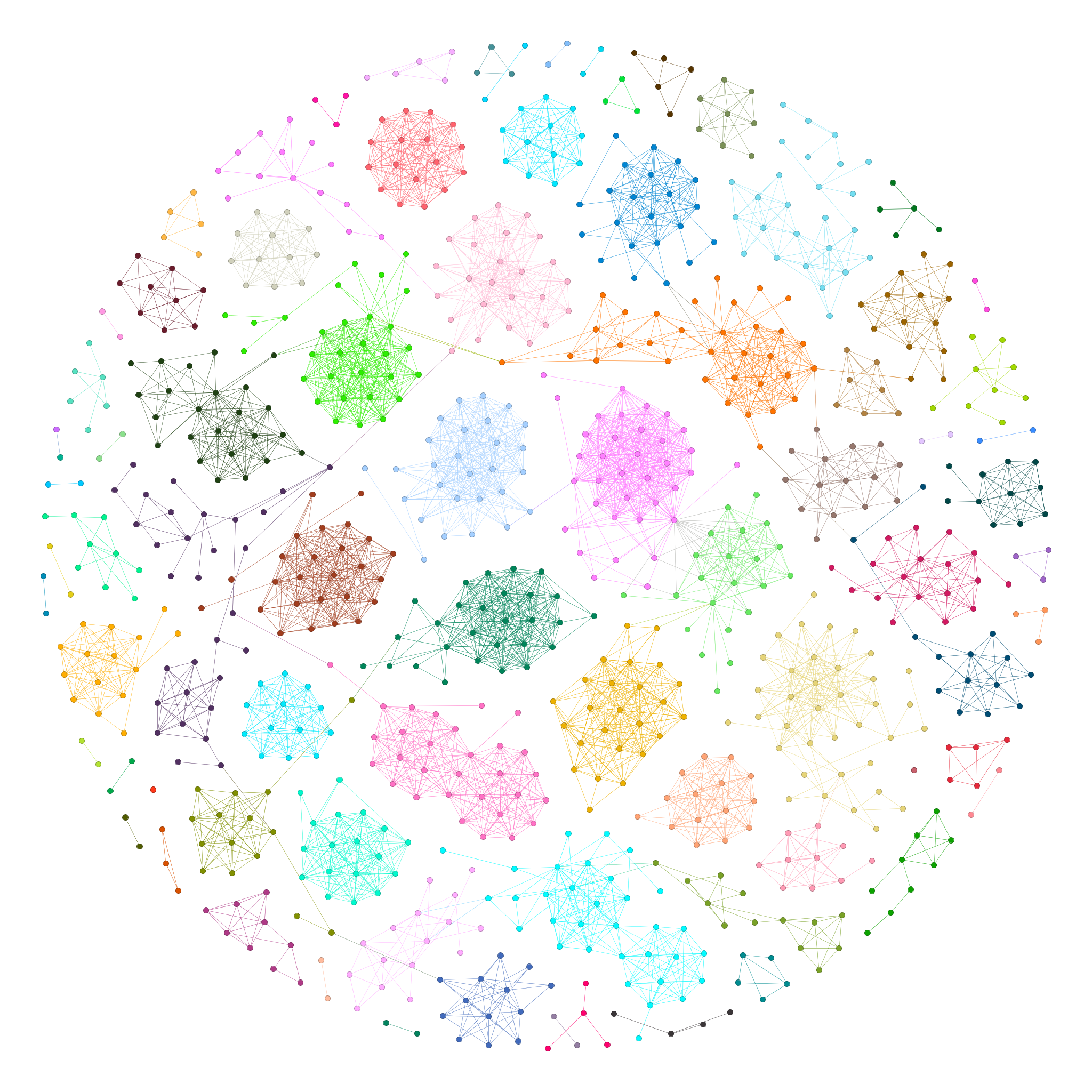} 
   \caption{Ensemble clustering of 875 economics journals 2006-2019. An edge between two nodes indicates that the pair is classified in the same community at least 80 percent of the time out of 3,000 repetitions of the Louvain algorithm. Unclassified journals (217) are not drawn. }
   \label{fig:ensemble_full}
\end{figure}

The stability of the structure of the three fused networks over time has suggested developing a unique classification of journals for all periods considered. The use of Louvain algorithm with an iterative ensemble clustering approach \citep{strehl2002} has allowed to reach a classification for 875 journals out of 1092 (80\%). The list of journals and their classification is reported in Appendix, Section A.3. This classification resulted in a fragmented landscape composed of 77 journals groups, as drawn in Figure \ref{fig:ensemble_full}. Among them, 34 include 10 or more journals, while 29 are composed of less than 5 journals each. The largest group has 37 journals. 

\begin{figure}[htbp]
    \centering

    \begin{adjustbox}{max totalheight = \textheight}
    \begin{floatrow}

        \ffigbox[\textwidth]{%
            \begin{minipage}{0.48\textwidth}
                \includegraphics[width=\linewidth]{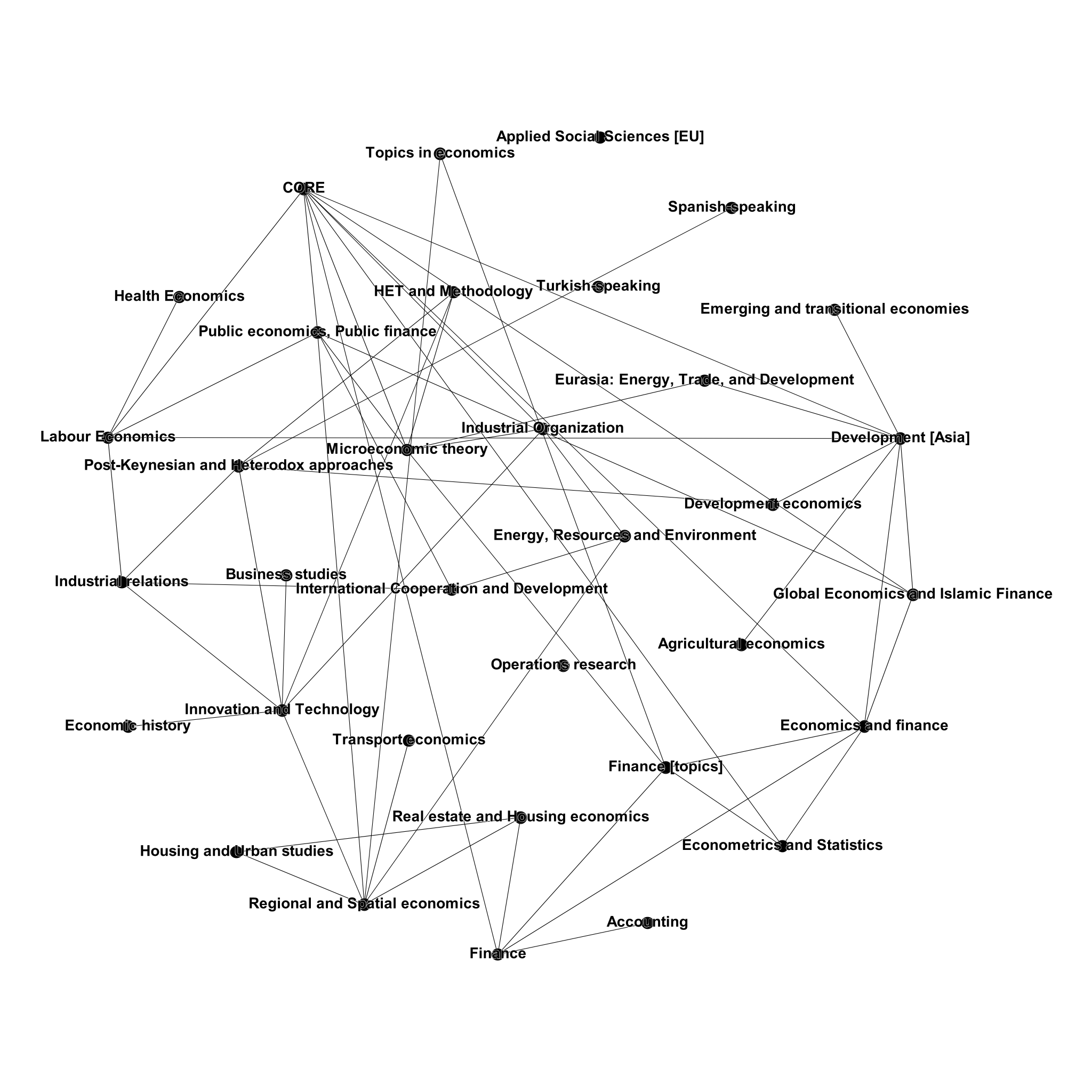}
                \centering (a) Editors
            \end{minipage}
            \hfill
            \begin{minipage}{0.48\textwidth}
                \includegraphics[width=\linewidth]{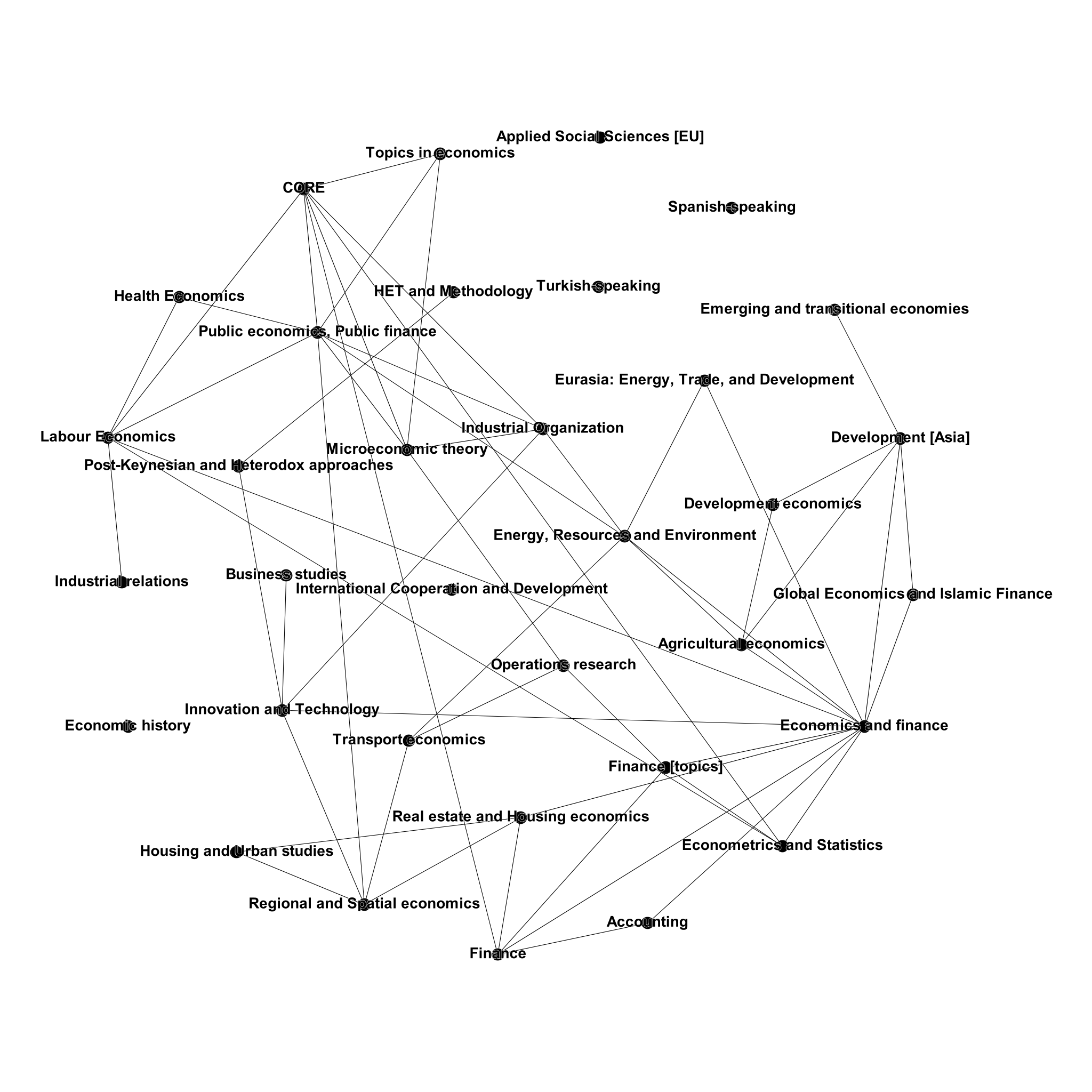}
                \centering (b) Authors
            \end{minipage}

            \vspace{0.3cm}

            \begin{minipage}{0.48\textwidth}
                \includegraphics[width=\linewidth]{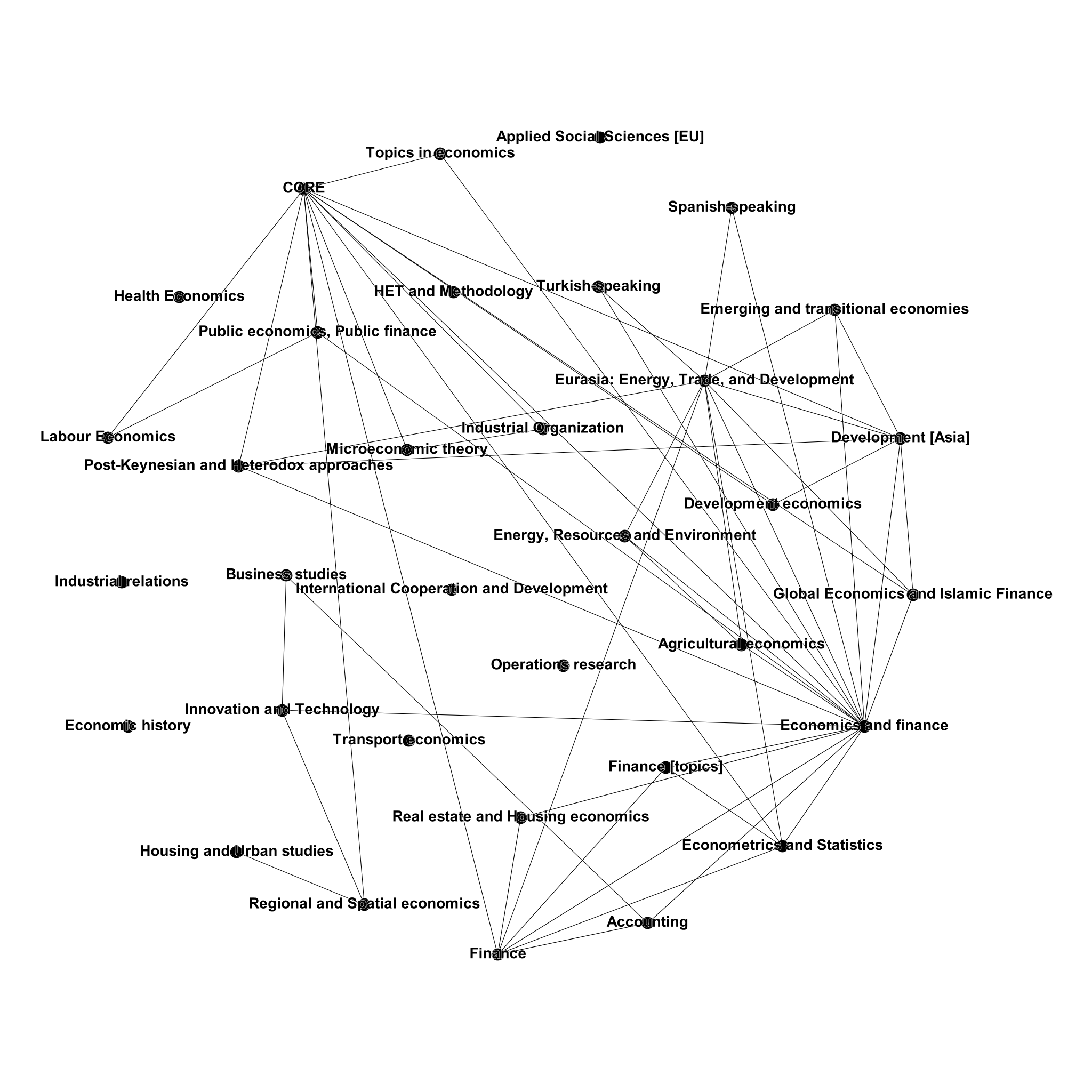}
                \centering (c) References
            \end{minipage}
            \hfill
            \begin{minipage}{0.48\textwidth}
                \includegraphics[width=\linewidth]{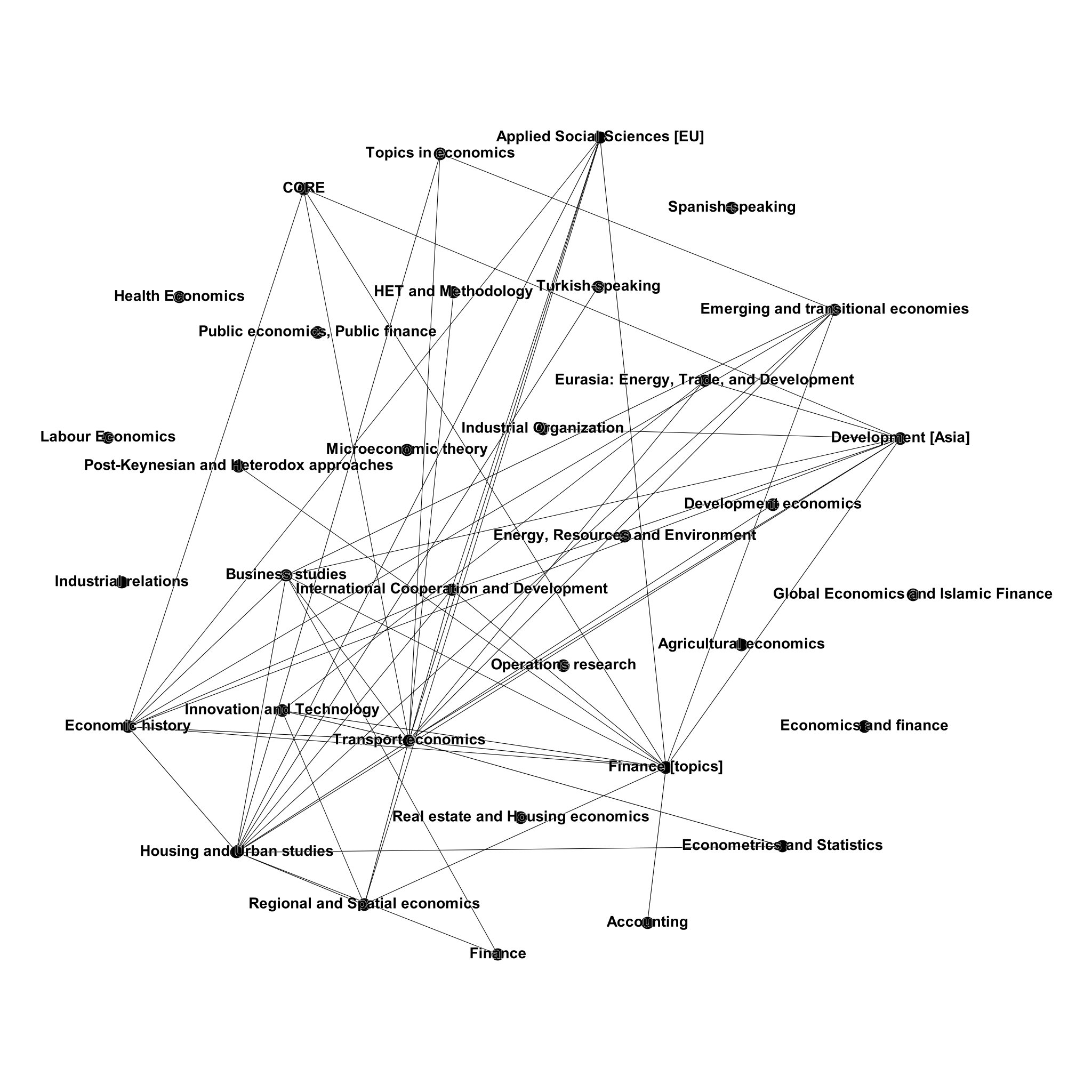}
                \centering (d) Abstracts
            \end{minipage}

            \vspace{0.3cm}

            \begin{minipage}{0.48\textwidth}
                \includegraphics[width=\linewidth]{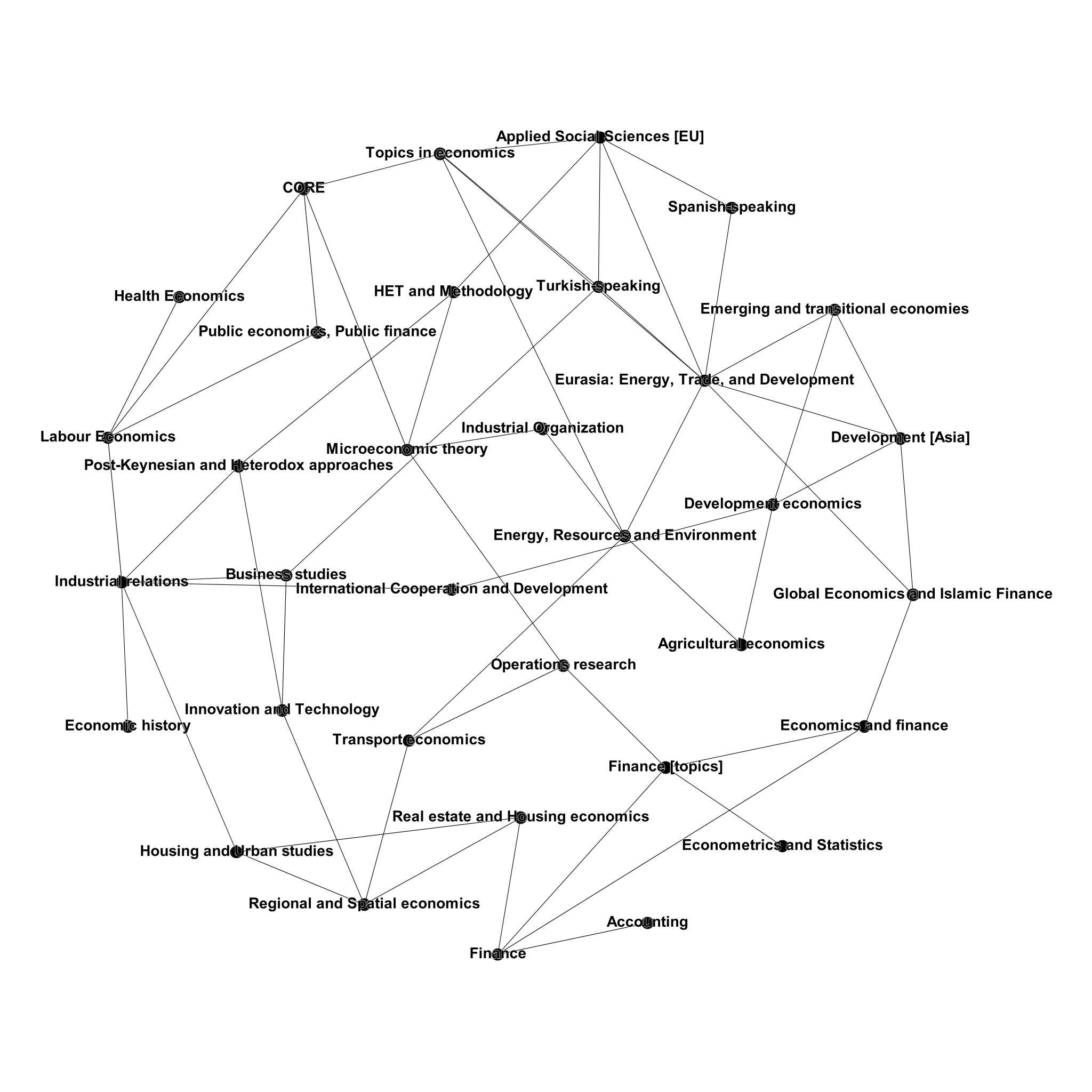}
                \centering (e) Fused
            \end{minipage}
            \hfill
            \begin{minipage}{0.48\textwidth}
                The panels of this figure were obtained by averaging entries of the similarity matrices to aggregate the communities of journals and visualize their relations. Each figure shows only the 10\% strongest edges. The layout of the network is fixed and given by the fused matrix.
                \captionof{figure}{Network visualization of the relationships among communities of journals}
                \label{fig:agg_networks}
            \end{minipage}
        }{}
    \end{floatrow}
    \end{adjustbox}
\end{figure}

Table \ref{tab:communities} reports the labels and the number of journals for the groups with more than 9 journals. The similarity relationships between groups of journals are drawn in the five panels of Figure \ref{fig:agg_networks}. Each panel draws a shrinked network where each group of journals is a node. The edges between groups are calculated as means of the similarities between the journals of the groups. The edges in the five panels of Figure \ref{fig:agg_networks} show the 10 percent strongest similarity between the journals groups in the fused matrix and in each of the four layers used to compute it. 

Appendix, Section A.4 reports for each group of journals the computation of PDCs between the fused matrix and each of the layers conditioned on the others. These data allow analyzing the contribution of the four layers of information to the internal structure of each group of journals in the fused network. In general, with some specific nuances, the internal structures of the groups seem to be driven mainly by social layers, especially that of editors. In most groups, the PDC values that refer to the layer of editors seem to increase over time. This result can be interpreted as a general trend toward strengthening the role of editors in structuring the discipline over time.

\begin{table}[]
\caption{Main communities of journals}
\small
\begin{tabular}{lc}
\toprule
\textbf{Community}                    & \textbf{Journals} \\ 
\midrule
Isolates (no community)                        & 217      \\ 
Smaller communities (\textless{}10 journals)   & 164      \\ 
Applied Social Sciences                        & 37       \\ 
Finance                                        & 36       \\ 
CORE                                           & 33       \\ 
Post-Keynesian and Heterodox approaches        & 33       \\ 
Global Economics and Islamic Finance           & 31       \\ 
Agricultural economics                         & 30       \\ 
Econometrics and Statistics                    & 30       \\ 
Accounting                                     & 28       \\ 
Regional and Spatial economics                 & 28       \\ 
Emerging and transitional economies            & 26       \\ 
International Cooperation and Development      & 26       \\ 
Business studies                               & 25       \\ 
Topics in Finance                              & 25       \\ 
Economics of Energy, Resources and Environment & 24       \\ 
Economics and finance                          & 22       \\ 
Economics of Innovation and Technology         & 22       \\ 
Development economics                          & 18       \\ 
Operations research                            & 18       \\ 
Microeconomic theory                           & 17       \\ 
Labour Economics                               & 16       \\ 
Health Economics                               & 15       \\ 
Industrial Organization                        & 15       \\ 
Industrial relations                           & 15       \\ 
Topics in Economics                            & 15       \\ 
HET and Methodology                            & 14       \\ 
Housing and Urban studies                      & 14       \\ 
Spanish-language                               & 14       \\ 
Eurasia: Energy, Trade, and Development        & 13       \\ 
Economic history                               & 13       \\ 
Public economics, Public finance               & 13       \\ 
Real estate and Housing economics              & 12       \\ 
Transport economics                            & 12       \\ 
Development {[}Asia{]}                         & 11       \\ 
Turkish journals                               & 10       \\ 
\bottomrule
\end{tabular}
\label{tab:communities}
\end{table}

For many groups of journals the label is straightforward, for others it is more difficult to understand the logic of clustering and also to choose a label. Moreover, the rationale behind the formation of the groups appears somewhat heterogeneous. 

A good understanding of some groups requires recalling that the journals considered are those indexed by EconLit. In EconLit ``journals are selected for inclusion [...] on the basis of their peer-reviewed economic content, which must be substantial or of equal emphasis in interdisciplinary journals'' \citep{econlit}. Therefore, EconLit also includes highly selected sets of journals of contiguous but autonomous \textit{disciplines}. 

There is no clear consensus in the literature on the definition of ``discipline'' \citep{trowler2012disciplines} and the fine line between it, research fields, and specialties \citep{Law_specialties, good, Coccia}. It is therefore delicate to distinguish the nature of some of the groups detected through ensemble clustering.
However, a few of these groups collect journals of contiguous disciplines that are indexed in EconLit. The \emph{Accounting} group gathers journals mainly devoted to accounting, which has developed as an autonomous academic discipline since 1960 \citep{zimmerman}. Figure \ref{fig:agg_networks} shows that \emph{Accounting} has a high level of similarity in the fused network only with Finance.

The \emph{Operation research} community collects a few journals of a discipline developed at the boundaries of economics, applied mathematics, and engineering \citep{Vertinsky}. In Figure \ref{fig:agg_networks} \emph{Operations research} is linked by a high level of similarity with three other groups, but only due to the crossed presence of authors, since it appears isolated in the network based on editors, abstract and references.

EconLit also indexes a set of statistics journals that ensembles clustering groups with econometrics journals in a cluster labeled \emph{Econometrics and Statistics}. 

The debate is ongoing about whether finance should be considered a discipline separate from Economics or a specialty inside economics. \citet[p. 105]{Fourcade} described the evolution of finance as a discipline by highlighting its ``institutional rise as an intellectual powerhouse within economics'' to the ``establishment of a teaching base'' in business schools. In fact, the clustering has detected up to three groups of finance journals: \emph{Finance}, \emph{Topics in Finance} and \emph{Economics and finance}. The first group gathers journals considered, for example in the CNRS rating of economics journals \citep{cnrs}, top-tier in finance, such as the \textit{Journal of Finance}, the second group appears to gather mostly second-tier journals, and the third less technical journals explicitly recalling in their title the connection with economics. Figure \ref{fig:agg_networks} shows that these three groups have a high level of similarities in the fused network and in the layers of authors, editors, and references, thus suggesting that the social communities are interrelated and that they share a common knowledge base. 

To discuss the rationale of the other groups, it is useful to retrieve a suggestion from \citet[p. 276]{Law_specialties}, who proposed to distinguish between ``technique'', ``theory'' and ``subject matter'' specialties.
A technique-based or method-based specialty ``constitutes an interacting group of scientists, whose solidarity rests on the basis of shared scientific gadgetry and its development. [...] preferred subject matters are defined only indirectly in relation to the strongly held methodological standards'' \citep{Law_specialties}. Theory-based specialties are defined in terms of a shared formalism. Their members' standards are mainly concerned with theory and its development, and examples related to various devices and problems will arise from this central concern with theory. Finally, subject-matter specialties are defined for handling a particular topic or problem, and their members are those who work on it. They are trained to use a variety of techniques and theories, none of which may be preferred in general \citep{Law_specialties}. 

Many of the groups appear to be explainable with such a general definition of specialties. Several of the larger groups seem to cluster the journals according to traditional specialties or subfields of economics. Actually, it appears that some of the groups are defined by a combination of the three criteria suggested by Law.

Theoretical orientation appears to play a key role in the formation of a major group, which includes 33 journals associated with the post-Keynesian tradition and other heterodox approaches. It gathers journals explicitly open to ``heterodox contributions'' such as the \textit{Cambridge Journal of Economics} and journals working on Sraffian and Post-Keynesian approaches. When confronted with the list of heterodox journals maintained by Jakob Kapeller and Niklas Klann\footnote{\url{https://heterodoxnews.com/hed/journals.html}}, the cluster appears as a small subset of it. This is due to a couple of reasons. The first is that EconLit does not provide comprehensive coverage of all journals, especially those that host articles on critical economic policy aimed at the general public or readers from other disciplinary fields such as history, philosophy or political sciences. The second is that ensemble clustering assigns some heterodox journals from Kapeller and Klann's list to different and specific groups. 
In particular, many of the list are classified in the clusters of \textit{History of Economic Thought and Methodology} and of \textit{Economics of Innovation and Technology}. The first is surely a subject-based cluster, while the second is defined at the boundary between subject-based and method-based, since it gathers many journals adopting an evolutionary approach to the study of economics in general. These three groups are linked by high levels of similarity in the fused network and are relatively isolated from the rest. The three clusters refer to overlapping communities of editors and authors, but when references and abstracts are considered, the similarities among them are not high (see Figure \ref{Communities_in_layers}). It is worth to note that \textit{Post-keynesian and Heterodox Approaches} share its knowledge base with the \textit{CORE}. This is very likely due to an asymmetric relationships between the two literature: while heterodox scholar discuss the literature produced by the \textit{CORE}, this latter simply ignore literature produced by heterodox scholars. This asymmetric relationship is for example documented in the case of DSGE and macro agent-based models \citep{crossfertil}.

The \textit{Microeconomic Theory} cluster can be interpreted in terms of a ``theory-based'' specialty. It gathers all relevant journals about microeconomic theory and mathematical economics; it also includes the journal published by the American Economic Association and devoted to microeconomics. It is strongly linked to \textit{CORE} through overlapped social communities of editors and authors, and share with it its knowledge base.

Another clear example of community interpretable in Law's framework is \textit{Econometrics and Statistics}, with a strong focus of methods-based specialization.

Almost all the other groups can be read as subject matter specialties or, more specifically, as applied subject matter specialties. They tend to bring together journals focused on specific areas of applied economics, including agricultural economics, energy and environmental economics, real estate and housing, the non-profit and cooperative sectors, transport economics, and economic history. Many groups have in their label the word ``development''. In these cases, the groups consist of journals addressing applied economic issues related to particular countries or macro-regions, such as those focused on Asia-Eurasia, the development of Asian economies, China, or, again, Global Economics and Islamic Finance. \citet{Kanbur} claims that there are many reasons to think that the distinction between economics and development economics ``will wither on the vine''. According to the data here considered, it appears that development economics is not a unitary subdiscipline of economics, but it appears fragmented in many different social and intellectual communities, with various degrees of overlap.

However, not all groups nicely align with Law's framework. A prime example is the group labeled \emph{CORE} with its 33 journals.

The label ``CORE'' is drawn from \citet{Craneinvis}. According to her, the disciplinary literature is ``fortunately [...] located in a few core journals'' that provide ``a kind of repetition in scientific communication insuring that certain ideas will be repeated and emphasized sufficiently so that the scientists who are interested in these problems will be sure of receiving at least some of the currently important messages and therefore continue to do research on these problems'' \citep[p. 113-114]{Craneinvis} and with these methods. The \textit{Journal of Economic Literature} and the \textit{Journal of Economic Perspective} fit perfectly with this description.

The CORE group gathers all the so-called Top Five journals of economics, five journals published by the American Economic Association, other general-interest journals such as the \textit{Economic Journal} and \textit{European Economic Review}, and many journals devoted to macroeconomics. Moreover, it contains also a few top-tier journals in a few specialized subdisciplines, specifically labor economics (\textit{Journal of Labor Economics)} and development economics (\textit{e.g. Journal of Development Economics}. This broad spanning and heterogeneous nature is at odds not only with Law's categories but with the idea of specialization itself. What is then the unifying factor of this group of journals?

A first clue can be found in Appendix, Section A.4: the CORE group has the strongest average value of PDC for the layer of editors, and their role in structuring the group appears to be increasing slightly over time. 
In Figure \ref{fig:agg_networks} it appears that, when the fused matrix is observed, the \emph{CORE} has strong similarities with journals of three other groups only: \textit{Public Economics and Public Finance}, \textit{Labour Economics} and \textit{Topics in Economics}. The relational structure of the network of editors and authors is completely different: the \textit{CORE} appears to have many strong similarity links with many clusters, among which \textit{Finance}, \textit{Econometrics and Statistics}, \textit{Microeconomic Theory}, \textit{Public Economics} and \textit{Labour Economics}. When references are analyzed, the high level of similarity between \textit{CORE}'s knowledge base and that of other clusters appears to be even more widespread. In contrast, when abstracts are examined, similarities seem limited to four specialties.

\begin{figure}
\centering
     \includegraphics [scale=0.23]{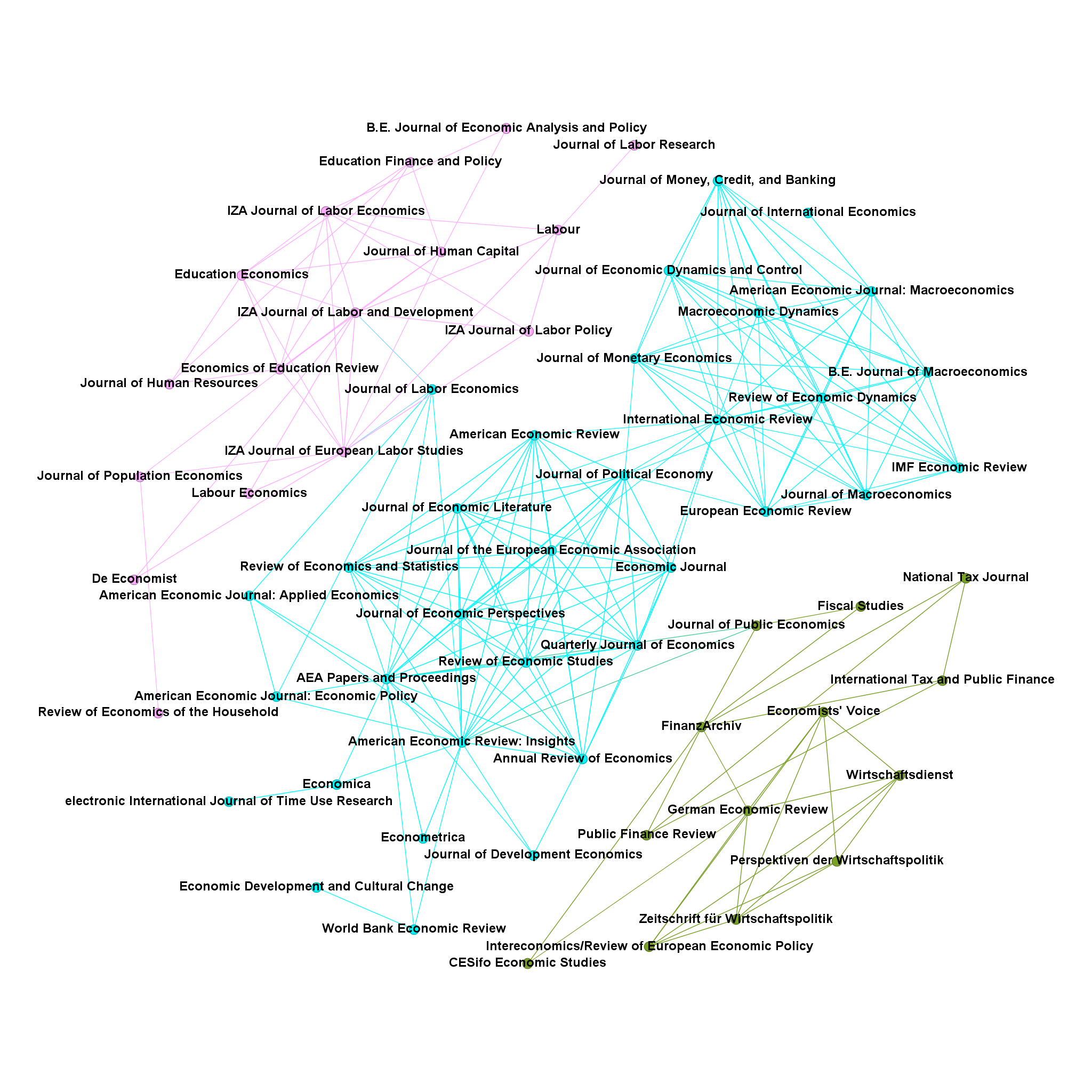} 
   \caption{Focus from Figure \ref{fig:ensemble_full} - Connected component including the \textit{Core}, \textit{Public economics} and \textit{Labour economics} communities.}
   \label{fig:ensemble_core}
\end{figure}

When the network of journals generated by the ensemble clustering is explored by searching the components of the network, the \textit{CORE} appears connected to \textit{Labour Economics}, \textit{Public Economic} and \textit{Public Finance} groups. Figure \ref{fig:ensemble_core} shows the links among the journals of the three clusters: a link indicates that a pair of journals is clustered together at least 80\% of the 3,000 repetitions of the Louvain algorithm. The structure of the network shows clearly that the CORE is in fact structured in two subgroups, one of general interest and the other of macroeconomic journals. The \textit{Journal of Political Economy} is the main bridge between the two subgroups. The \textit{Journal of Labor Economics} creates a link with the \textit{Labour Economics} cluster, by being connected with all IZA journals. Finally, the \textit{Journal of Public Economics} is the bridge between the \textit{CORE} and the \textit{Public Economics and Public Finance} group.

All of this suggests that the unifying forces of this group are not dynamics of specialization, but rather social relations of prestige and hierarchy. The centrality of the \textit{CORE} when it comes to the editors network speaks of a mechanism of co-opting and/or mobilizing influential figures from disciplinary specialties to a central microcosm aiming at mirroring the entire discipline. Individual specialties, in turn, are legitimized by this process and gain prestige along with the people embodying these links.

As a final note, many of the smaller groups appear to gather journals according to the country of publication, again detaching them from specialization as their unifying mechanism. Examples include journals from Turkey, France, Slovenia, Croatia, Brazil, Spain, and South Africa. This highlights the existence of probably long-standing national communities of economists \citep{dippoliti}.

\section {Discussion}

Economics journals were chosen as the basic unit of exploratory analysis because they represent institutionalized spaces in which the social and intellectual relations that structure a discipline can be observed. This is even more true when, as in economics, journals represent the main outlet for the publication of research findings. Two social communities are gathered around journals: editors acting as gatekeepers of science and authors publishing their ideas. From an intellectual standpoint, journals can be conceived as curated collections of articles communicating ideas. These articles are built by using a knowledge base that is collected in their bibliographies. 

The relationships between journals can represented as a four-layer network, where each layer contains information related to editors, authors, abstracts, and cited references. This paper has developed quantitative techniques to define similarity relationships between journals based on these four layers of information. Social similarities between journals are measured for editors and authors by leveraging on, respectively, the phenomena of interlocking editorship and interlocking authorship. Intellectual similarities between journals are defined in terms of the distance between vectors embedding the abstracts of the published articles. Similarities between the knowledge bases of journals is measured by leveraging on bibliographic coupling. These four layers of social and intellectual similarities are finally integrated in a unique fused similarity network of journals using a suitable algorithm (SNF). This exploratory framework was adapted to handle the evolution of journal relationships in three periods, one before and two after the financial crisis, with the aim of documenting whether and how this external shock impacted economics as a discipline. 

Some results of the analysis emerge from the examination of the three periods separately. The first corroborates previous studies \citep{baccini_intellectual, baccini_sim_net_fus}, indicating that social communities formed around journals are highly interconnected; specifically, the relationships between journals established by editors and authors overlap significantly. Furthermore, these relationships are strongly correlated with those identified through bibliographic coupling. In contrast, relationships based on content similarity tend to exhibit weaker associations with the others. This last result allows the generalization of similar findings previously obtained from the analysis of articles within a single economics journal \citep{baccini_cje}. It suggests that relying solely on textual analysis may provide an incomplete representation of the research landscape across various levels of granularity. 

A second result, also confirming previous findings \citep{baccini_sim_net_fus, baccini_cje}, is that when similarity relations between journals are observed separately for each individual layer of information, community detection algorithms do not produce robust classifications. When, instead, more layers of information are integrated through the SNF in a unique similarity network, community detection algorithms produce strong classifications of journals. 

A third result generalizes some previous specific findings by showing that the structure of the relationships between journals as represented in the fused networks depends for all three periods mainly on the editors \citep{baccini_sim_net_fus}. This suggests the following conjecture: the editors select the papers, and as a result the user community of each individual journal. This makes the relationships between journals largely dependent on the action of the editors as gatekeepers of science. Their authority in deciding about the articles to be published defines the boundaries of the communities of authors gathered around journals, and also of their contents and knowledge base. 

One possible objection to this line of reasoning is that we have considered full editorial boards of journals when the decision-making power on publication is primarily in the hands of a subset of board members who have the highest editorial ranking in the journal. The counter-objection is based on previous research \citep{baccini_re} showing that the relational structure between economics journals is the same when looking at entire editorial board, and when considering considering only subsets of carefully individuated editorial leaders. In fact, this second network is simply a ``distillation'' of the first, reproducing the same relationship only more sparsely. The use of entire editorial boards has the main advantage of avoiding the difficult and tedious task of classifying certain board members as editorial leaders.

The role of editors as gatekeepers of science may be the main channel to build and strengthen hierarchy, consensus, and control that, according to \citet[p. 96]{Fourcade}, distinguish economics from the other social sciences. They suggested that ``those who control the top journals'' promote consensus by defining ``what constitutes quality research'' and ``conversely, control might be more effective and enforceable because there is more consensus''. 
The results of our exploratory analysis may suggest that this mechanism is not limited to the top journals, but it is active in many of the journal clusters we identified. It is a matter of future inquiries whether this mechanism is peculiar of economics with respect to other disciplines, and whether it has played a role in the unusual insularity of economics as documented by \citet{Truc_2023}.

Other results of the analysis came from the comparison of the relationships between journals in the three periods considered. The main regards the discovery of a stable structure of relationships among journals in the years before and after the financial crisis: the structure of the fused networks remains completely unchanged between 2006 and 2020. It indicates that the social and intellectual relationships between economics journals are completely stable in the three periods analyzed. This stability suggested adopting a new strategy for detecting clusters of journals leveraging the classical Louvain algorithm, but now coupled with an ensemble clustering approach: two journals are linked together only if they co-occurred in the same cluster in at least 80\% of the algorithm's runs across the three periods considered. This procedure has generated a unique and strong clustering of journals for the whole period by adding stability to an overall stable landscape. 

This stable relational structure resulted in a fragmented landscape where about 80 percent of economics journals belong to 74 clusters, 34 of which are composed of more than 9 journals. Some of these groups are made up of journals from contiguous disciplines, such as accounting or operations research. Others are journals that share a common theoretical orientation, such as heterodox and post-Keynesian economics, or the economics of innovation and technology; and still others are composed of techniques such as statistics and econometrics. Most groups are defined as applied subject specialties, collecting journals devoted to specific areas of applied economics such as agriculture, real estate, transport, and economic history. Finally, the CORE group includes the so-called Top Five journals of economics, several AEA journals, general-interest European journals, and top field journals in macroeconomics, labor, and development. Its coherence is not based on disciplinary specialization, but on editorial prestige and hierarchical social ties. Structurally, it splits into subgroups of general interest and macroeconomics, connected through key bridging journals. The CORE functions as the center of the discipline co-opting influential scholars across specialties into a shared academic elite. 

Perhaps the most accurate image of the discipline is that of a highly fragmented field, where both the so-called mainstream and heterodox traditions exhibit considerable internal diversity \citep{dippoliti}. Our contribution to the ongoing debate is to show that this fragmentation, typically seen as the outcome of intellectual disputes or a ``battle of ideas'', is, in fact, primarily shaped by the segmentation of scholarly communities, most notably among editors, clustered around journals.

This fragmented landscape of journals appears to be largely stable before and after the financial crises. Similarly, the relational structure between journals as a whole and within each cluster also appears to be largely stable. Stability should be interpreted differently for social and intellectual relations. 

During a period of more than 10 years, each journal underwent a partial change in both editorial board members and contributing authors. Nevertheless, the social structure linking journals, driven by interlocking editorship and authorship, remained largely stable. This suggests that both editorial boards and authors tend to reproduce established patterns of association despite individual-level changes. As noted in \citet{baccini_re}, “links between pairs of journals tend to be redundant”, a charachteristic that can be interpreted through the lens of social and intellectual homophily, both within individual boards and across journals belonging to the same group. The persistence of these ties over time indicates that editorial turnover tends to preserve existing inter-journal connections; the redundancy of these links likely underpins the overall stability of the network of editorial relationships.
A similar mechanism appears to govern interlocking authorship. Author communities create redundant ties across journals, and editorial selection processes tend to reinforce the same relational structures over time. Together, these dynamics contribute to the long-term continuity of the social structure of the journals network.

Analogously, the stability of intellectual relationships between journals is not necessarily the result of a discipline that avoids shifting topics. As a response to the challenge represented by the financial crisis, economists switched their focus to studying its causes and consequences \citep{LEVY2022}. This aligns with the normal condition of problem-solving: the ability of a discipline to redirect attention to emergent puzzles without inherently altering its worldview or methodology. In this sense, intellectual continuity coexists with responsiveness to new problems and neither presupposes a radical transformation of the foundations of the fields \citep{mirowski}. For example, \citet{aignerkapeller} documented that the increased attention to financial instability ``did not lead to any major theoretical or methodological changes in contemporary economics''. 

It can even be conjectured that the financial crisis created a period of cognitive uncertainty that fueled a reaction of ``increasing defensiveness on the part of different subgroups, concerning their own interpretation of the intellectual problems'' \citep[p. 37]{Craneinvis}. \citet{mirowski} refereed explicitly to the defensive strategy of neoclassical economics in the aftermath of the crisis when ``intrepid intrepid economists were busy gestating all sorts of `new economic theory to explain the crisis, when in fact they had mostly dug themselves into a
defensive crouch where nothing about the orthodoxy was being sloughed
off, and no errors would be freely conceded''. This may have produced increasing differentiation between groups and a decline in the exchange of ideas. 

Finally, the generalizability of these results to other disciplinary areas remains an open question. On the one hand, empirical comparative accounts \citep[e.g.][]{Fourcade} point in the direction of a distinctiveness in the nature of economics when compared to other social sciences in terms of, for example, hierarchical organization. On the other hand, putting the emphasis only on the peculiarities of economics as a discipline risks falling into a form of ``economics exceptionalism'' of sorts. In the end, this is an empirical question that only more comparative research will be able to solve.

\section{Conclusion}

The analysis developed in this paper suggests that any changes in research topics or modifications to the knowledge base that may have occurred during the 15-year observation period appear not to have altered fundamental social and intellectual relationships among journals. By extension, the discipline's institutional structure demonstrates remarkable stability.
This finding underscores the importance of distinguishing between what economists study and how their work is organized, evaluated, and legitimized. Although what economists study can be shaped by external crises, policy demands, or methodological innovations, the institutional dimensions of their work are governed by institutionalized norms and power relations. These norms and power relations are mainly defined in social communities gathered around journals, a structure that may well characterize other disciplines, though this remains to be tested.

The network of journals defines the relational infrastructure of economics where editors operate as gatekeepers. They reinforce shared evaluative criteria and reproduce hierarchies, even as the discipline absorbs new topics. In this context, the crisis was a new puzzle to solve within the existing paradigm, not a challenge to the authority structures governing publication or collaboration. The financial crisis has not destabilized the social order of the discipline. 

That social order is built in such a way that the costs of defection from dominant norms outweigh the benefits of radical innovation. When new approaches or themes, such as behavioral or experimental economics, are introduced, they must conform to institutionalized standards of validity by passing through established social and intellectual channels, most notably publication in top-tier journals. Disruptive ideas, particularly heterodox critiques, tend to appear in journals that do not contribute to the knowledge base of the CORE. As a result, they struggle to take root and are at best merely tolerated or at worst dismissed as fringe science or pseudoscience. \citet{Chari} captured this logic with striking clarity: ``The recent crisis has raised, correctly, the question of how best to improve modern macroeconomic theory. I have argued we need more of it. After all, when the AIDS crisis hit, we did not turn over medical research to acupuncturists''. The economics profession has not seen the crisis as an opportunity to reconsider institutional structures of the discipline or its foundational assumptions, but rather as a technical challenge to be resolved within existing social and intellectual settings. It reflects a broader mechanism of epistemic closure, in which authority is reinforced by boundary work of gatekeepers who define what counts as legitimate knowledge and who is entitled to produce it. Whether other disciplines would exhibit similar mechanisms of epistemic closure and social stability when confronted with external shocks remains an open empirical question that the multilayer network framework developed in this paper is well equipped to address.

\section*{Declarations}

\begin{itemize}
\item Funding: the research is funded by the Italian Ministry of University, PRIN project: How economics is changing: A multilayer network analysis of the recent evolution of economics journals, between specialization and self-referentiality (1980-2020),  2022SNTEFP, PI: Alberto Baccini.
\item Conflict of interest/Competing interests: The authors have no competing interests to declare that are relevant to the content of this article.
\end{itemize}

\bibliography{references}  

\





\newpage

\section*{Appendix}

\subsection*{A.1. Transformation of cosine similarity}

Assuming that $x$ and $y$ are two vectors in $\mathbb{R}^d$, the cosine similarity is commonly defined as $C(x,y)=\frac{\langle x,y\rangle}{||x||\cdot||y||}$, where ${||\cdot||}$ and $\langle\cdot,\cdot\rangle$ respectively represent the usual vector norm and inner product. However, $C(x,y)$ can eventually be negative and in addition the cosine distance $1-C(x,y)$ is not a proper metric since it does not exhibit the triangle inequality property. Thus, an appropriate trasformation of cosine similarity is given by
    \begin{equation*}
        S(x,y) = 1 - \frac{1}{2}\sqrt{2(1 - C(x,y))},
    \end{equation*}
which produces an invariably positive similarity such that $1-S(x,y)$ is a proper metric.

\subsection*{A.2. Partial distance correlation with more than three similarity matrices} We
assume the notation given by \cite{szekely2014}. In particular, we also assume the definition of the distance correlation $R_{x,y}^\ast$ for two similarity matrices $x$ and $y$, as given in formula (3.8) by these authors. In this case, when three similarity matrices are considered, say $x$, $y$ and $z$, the partial distance correlation between $x$ and $y$ given $z$ is
given by
    \begin{equation*}
        R_{x,y\mid z}^\ast=\frac{R_{x,y}^\ast-R_{x,z}^\ast R_{y,z}^\ast}{\sqrt{1-(R_{x,z}^\ast)^2}\sqrt{1-(R_{y,z}^\ast)^2}},
    \end{equation*}
see formula (3.9) by \cite{szekely2014}. \cite{szekely2014} also introduce an algorithm for the practical computation of the previous formula.

In order to extend the partial distance correlation to the case involving more than three similarity matrices, let us assume that $s$ is a collection of similarity matrices, i.e. $s=\{s_1,\ldots,s_k\}$ with $k>3$. Therefore, we suggest a recursive procedure based on the formula 
    \begin{equation*}
        R_{s_i,s_j\mid s\backslash\{s_i,s_j\}}^\ast=\frac{R_{s_i,s_j\mid
        s\backslash\{s_i,s_j,s_k\}}^\ast-R_{s_i,s_k\mid
        s\backslash\{s_i,s_j,s_k\}}^\ast R_{s_j,s_k\mid
        s\backslash\{s_i,s_j,s_k\}}^\ast}{\sqrt{1-(R_{s_i,s_k\mid
        s\backslash\{s_i,s_j,s_k\}}^\ast)^2}\sqrt{1-(R_{s_j,s_k\mid
        s\backslash\{s_i,s_j,s_k\}}^\ast)^2}},
    \end{equation*}
which is actually a generalization of the usual recursive expression for computing partial correlations (see e.g. \citealt{kim}). 

In the simplest situation for $k=4$, we deal with four similarity matrices and adopt the notation $s=\{x,y,z,v\}$ for simplicity. For example, $x$ can represent the fused matrix, $y$ the editor similarity matrix, $z$ the abstract similarity matrix, and $v$ the reference similarity matrix. In this case, if $x$ is considered fixed, we have the three partial distance
correlations
    \begin{equation*}
        R_{x,y\mid z,v}^\ast=\frac{R_{x,y\mid z}^\ast-R_{x,v\mid z}^\ast R_{y,v\mid
        z}^\ast}{\sqrt{1-(R_{x,v\mid z}^\ast)^2}\sqrt{1-(R_{y,v\mid
        z}^\ast)^2}},\,
        R_{x,z\mid y,v}^\ast=\frac{R_{x,z\mid y}^\ast-R_{x,v\mid y}^\ast R_{z,v\mid
        y}^\ast}{\sqrt{1-(R_{x,v\mid y}^\ast)^2}\sqrt{1-(R_{z,v\mid y}^\ast)^2}}
    \end{equation*}
and
    \begin{equation*}
        R_{x,v\mid y,z}^\ast=\frac{R_{x,v\mid y}^\ast-R_{x,z\mid y}^\ast R_{v,z\mid
        y}^\ast}{\sqrt{1-(R_{x,z\mid y}^\ast)^2}\sqrt{1-(R_{v,z\mid y}^\ast)^2}}.
    \end{equation*}
On the basis of the definition and using the previous expressions for $k=5$, we can easily compute the partial distance correlation in the case of five similarity matrices assuming the notation $s=\{x,y,z,v,w\}$ for simplicity. If $x$ is considered in turn as fixed, we have
    \begin{equation*}
        R_{x,y\mid z,v,w}^\ast=\frac{R_{x,y\mid z,v}^\ast-R_{x,w\mid z,v}^\ast
        R_{y,w\mid z,v}^\ast}{\sqrt{1-(R_{x,w\mid z,v}^\ast)^2}\sqrt{1-(R_{y,w\mid
        z,v}^\ast)^2}},\,
        R_{x,z\mid y,v,w}^\ast=\frac{R_{x,z\mid y,v}^\ast-R_{x,w\mid y,v}^\ast
        R_{z,w\mid y,v}^\ast}{\sqrt{1-(R_{x,w\mid y,v}^\ast)^2}\sqrt{1-(R_{z,w\mid
        y,v}^\ast)^2}},
    \end{equation*}
    \begin{equation*}
        R_{x,v\mid y,z,w}^\ast=\frac{R_{x,v\mid y,z}^\ast-R_{x,w\mid y,z}^\ast
        R_{v,w\mid y,z}^\ast}{\sqrt{1-(R_{x,w\mid y,z}^\ast)^2}\sqrt{1-(R_{v,w\mid
        y,z}^\ast)^2}},\,
        R_{x,w\mid y,z,v}^\ast=\frac{R_{x,w\mid y,z}^\ast-R_{x,v\mid y,z}^\ast
        R_{w,v\mid y,z}^\ast}{\sqrt{1-(R_{x,v\mid y,z}^\ast)^2}\sqrt{1-(R_{w,v\mid
        y,z}^\ast)^2}},
    \end{equation*}
where 
    \begin{equation*}
        R_{x,w\mid z,v}^\ast=\frac{R_{x,w\mid z}^\ast-R_{x,v\mid z}^\ast R_{w,v\mid
        z}^\ast}{\sqrt{1-(R_{x,v\mid z}^\ast)^2}\sqrt{1-(R_{w,v\mid
        z}^\ast)^2}},\,
        R_{y,w\mid z,v}^\ast=\frac{R_{y,w\mid z}^\ast-R_{y,v\mid z}^\ast R_{w,v\mid
        z}^\ast}{\sqrt{1-(R_{y,v\mid z}^\ast)^2}\sqrt{1-(R_{w,v\mid
        z}^\ast)^2}},
    \end{equation*}
    \begin{equation*}
        R_{x,w\mid y,v}^\ast=\frac{R_{x,w\mid y}^\ast-R_{x,v\mid y}^\ast R_{w,v\mid
        y}^\ast}{\sqrt{1-(R_{x,v\mid y}^\ast)^2}\sqrt{1-(R_{w,v\mid
        y}^\ast)^2}},\,
        R_{z,w\mid y,v}^\ast=\frac{R_{z,w\mid y}^\ast-R_{z,v\mid y}^\ast R_{w,v\mid
        y}^\ast}{\sqrt{1-(R_{z,v\mid y}^\ast)^2}\sqrt{1-(R_{w,v\mid
        y}^\ast)^2}},
    \end{equation*}
    \begin{equation*}
        R_{x,w\mid y,z}^\ast=\frac{R_{x,w\mid y}^\ast-R_{x,z\mid y}^\ast R_{w,z\mid
        y}^\ast}{\sqrt{1-(R_{x,z\mid y}^\ast)^2}\sqrt{1-(R_{w,z\mid
        y}^\ast)^2}},\,
        R_{v,w\mid y,z}^\ast=\frac{R_{v,w\mid y}^\ast-R_{v,z\mid y}^\ast R_{w,z\mid
        y}^\ast}{\sqrt{1-(R_{v,z\mid y}^\ast)^2}\sqrt{1-(R_{w,z\mid
        y}^\ast)^2}},
    \end{equation*}
    \begin{equation*}
        R_{x,w\mid y,z}^\ast=\frac{R_{x,w\mid y}^\ast-R_{x,z\mid y}^\ast R_{w,z\mid
        y}^\ast}{\sqrt{1-(R_{x,z\mid y}^\ast)^2}\sqrt{1-(R_{w,z\mid
        y}^\ast)^2}},\,
        R_{v,w\mid y,z}^\ast=\frac{R_{v,w\mid y}^\ast-R_{v,z\mid y}^\ast R_{w,z\mid
        y}^\ast}{\sqrt{1-(R_{v,z\mid y}^\ast)^2}\sqrt{1-(R_{w,z\mid
        y}^\ast)^2}}.
    \end{equation*}
Further extensions to the case $k>5$ are straightforward, although at the cost of increasing notational and computational complexity.

\newpage
\subsection*{A.3. List of journals and their classification in communities}

\newcolumntype{P}[1]{>{\raggedright\arraybackslash}p{#1}}
{\small
\begin{longtable}{P{8cm}P{7cm}}

\toprule
\textbf{Journal (Title EconLit)} & \textbf{Cluster} \\
\midrule
\endfirsthead

\toprule
\textbf{Journal (Title EconLit)} & \textbf{Cluster} \\
\midrule
\endhead

International Journal of Housing Markets and Analysis&Real estate and Housing economics\\
Journal of European Real Estate Research&Real estate and Housing economics\\
Journal of Housing Economics&Real estate and Housing economics\\
Journal of Housing Research&Real estate and Housing economics\\
Journal of Property Research&Real estate and Housing economics\\
Journal of Real Estate Finance and Economics&Real estate and Housing economics\\
Journal of Real Estate Literature&Real estate and Housing economics\\
Journal of Real Estate Portfolio Management&Real estate and Housing economics\\
\bottomrule
\end{longtable}
}

\newpage
\subsection*{A.4. Partial distance correlation in journal communities}

\newcolumntype{C}[1]{>{\centering\arraybackslash}p{#1}}

\begin{center}
\begin{longtable}{ll@{} l *{4}{C{2cm}} @{}}
\toprule
\textbf{PDC} & \textbf{average} & \textbf{2006} & \textbf{2012} & \textbf{2019} \\
\midrule
\endfirsthead

\toprule
\textbf{PDC} & \textbf{average} & \textbf{2006} & \textbf{2012} & \textbf{2019} \\
\midrule
\endhead

\midrule
\multicolumn{5}{r}{\textit{Continued on next page}} \\
\midrule
\endfoot

\bottomrule
\endlastfoot

\multicolumn{5}{c} {\textbf{ACCOUNTING}} \\ 
Fused-Abstracts & 0.346 & 0.305 & 0.2 & 0.515 \\ 
 Fused-Editors & 0.394 & 0.423 & 0.261 & 0.491 \\ 
  Fused-Authors & 0.017 & -0.099 & 0.198 & 0.176 \\ 
  Fused-References & 0.499 & 0.554 & 0.401 & 0.501 \\  
\midrule
\multicolumn{5}{c}{\textbf{AGRICULTURAL ECONOMICS}}\\
Fused-Abstracts & 0.457 & 0.243 & 0.466 & 0.345 \\ 
  Fused-Editors & 0.706 & 0.605 & 0.594 & 0.704 \\
\end{longtable}
\end{center}

\end{document}